\documentclass[12pt,a4paper]{article}
\usepackage[T1]{fontenc}
\usepackage[margin=2.4cm]{geometry}
\usepackage{amsmath,amssymb}
\usepackage{graphicx}
\usepackage{bm}
\usepackage{booktabs}
\usepackage[round,authoryear]{natbib}
\bibpunct{(}{)}{;}{a}{}{,}
\usepackage{setspace}
\usepackage[colorlinks=true,linkcolor=blue,citecolor=blue,urlcolor=blue]{hyperref}
\numberwithin{equation}{section}
\newcommand{\bcdot}{\boldsymbol{\cdot}}
\newcommand{\bnabla}{\boldsymbol{\nabla}}
\newcommand{\PT}{\mathbb{P}}
\newcommand{\Uc}{\boldsymbol{\mathcal{U}}}

\begin{document}

\title{Iterative Construction of $n$-Dimensional Navier--Stokes Solutions
with Non-Gradient Convection}

\author{R. K. Michael Thambynayagam\thanks{The work reported here was carried out
independently by the author after retirement; the affiliation shown is his former
position.}\\[4pt]
\normalsize Managing Director (Retired), Schlumberger Cambridge Research\\[4pt]
\normalsize E-mail: \texttt{michael.thambynayagam@gmail.com}}

\date{\today}

\maketitle

\begin{abstract}
Exact periodic solutions of the incompressible Navier--Stokes equations arise when the convective
field is a pure gradient and can be absorbed into the pressure. We construct solutions outside this
class, in arbitrary spatial dimension, by an iteration in which each step solves a linear diffusion
problem forced by the transverse part of the preceding convective field. The construction is
dimension-independent, and for a cyclic family of initial fields the first two iterates are
available in closed form.

Carried out numerically in three dimensions, the iteration converges: at $\mathcal{R}e=10$
successive iterates contract with an observed factor $0.43$ and the limit is a fixed point of the
discretised map to a relative defect of $8.2\times10^{-5}$. The observed factor exceeds unity by
$\mathcal{R}e=30$, which identifies the practical range of the iteration rather than a proved
convergence boundary. We also show what goes wrong when the projection is omitted, as it was in an
earlier iteration of the author's: the iterates cease to be divergence-free and an apparent growth
appears that belongs to the expansion and not to the flow. For an $n$-dimensional generalisation of
the initial field the fraction of the convective field not absorbed by pressure is exactly
$2\sqrt{2}/3$ in every dimension $n\ge3$, and the longitudinal fraction exactly $1/3$. The
divergence of each returned source is a sharp diagnostic, failing by sixteen orders of magnitude at
the step where the projection is omitted; the cell kinetic energy is a weaker one, detecting gross
failure only.
\end{abstract}



\section{Introduction}
\label{sec:intro}

Closed-form solutions of the incompressible Navier--Stokes equations are rare, and known examples
are typically obtained by arranging for the nonlinear term to vanish from the velocity dynamics. In
particular, if the convective field $\boldsymbol{g}=(\boldsymbol{v}\bcdot\bnabla)\boldsymbol{v}$ is
a pure gradient, it can be absorbed entirely into the pressure, leaving the velocity to satisfy a
linear diffusion equation. The Taylor vortex \citep{taylor1923} is a classical example, as are the
periodic families classified by \citet{ant2020}, \citet{tha2023} and \citet{tha2026}. This
mechanism, however, applies only to a restricted class of flows and is unavailable for a general
solenoidal initial field.

In this paper we construct solutions beyond this gradient class in arbitrary spatial dimension.
Rather than requiring the convective field to be absorbed entirely into the pressure, we retain its
non-gradient component and carry it as a source, generating it iteratively: each step is a linear
diffusion problem whose forcing is the transverse part of the convective field computed at the step
before. The first two steps are available in closed form, and the construction is
dimension-independent.

\subsection{Iterative constructions: a brief review}
\label{sec:lit}

The iteration employed here is the classical mild-solution construction. \citet{fk1964}
reformulated the Navier--Stokes initial-value problem as an integral equation in which the semigroup
generated by the Stokes operator acts on the projected nonlinear term, and established existence
through a contraction argument for sufficiently small data or sufficiently short times.
\citet{kato1984} developed this approach in the $L^p$ setting that has since become standard,
obtaining strong solutions in $\mathbb{R}^m$ and clarifying their relation to weak solutions. For
the present work the relevant point is structural rather than analytical: at each stage the
nonlinear source is projected onto its solenoidal component before being propagated by the Stokes
semigroup, while convergence of the iteration is guaranteed only under appropriate bounds on the
data and time interval.

The purpose of those constructions is primarily to establish existence and regularity, rather than
to evaluate the iteration explicitly. Our objective is different: we carry out the iteration for a
concrete initial field, obtain explicit expressions for the resulting velocity and pressure, and
determine the parameter range over which the construction remains applicable. Expansions about the
Stokes solution are familiar in low-Reynolds-number fluid mechanics, but, to our knowledge, an
explicit iteration of the present form---with the solenoidal projection evaluated exactly at each
stage---has not been developed for periodic fields in arbitrary spatial dimension.

\subsection{Statement of the problem}
\label{sec:problem}

On the periodic cell, with no applied force,
\begin{equation}
\label{eq:ns}
\frac{\partial\boldsymbol{v}}{\partial t}
= \kappa\Delta\boldsymbol{v} - \bnabla p/\rho - \boldsymbol{g},
\qquad
\bnabla\bcdot\boldsymbol{v}=0,
\qquad
\boldsymbol{g}=(\boldsymbol{v}\bcdot\bnabla)\boldsymbol{v},
\end{equation}
with $\boldsymbol{v}(\boldsymbol{x},0)=\boldsymbol{v}^0$ smooth and solenoidal.

\citet{tha2023} recast \eqref{eq:ns} into three contributions---viscous, inertial and
applied---and showed that when the inertial contribution satisfies
\begin{equation}
\label{eq:cond}
g_i(\boldsymbol{x},t) = \frac{\Gamma(n/2)}{2\pi^{n/2}}
\int_{\mathbb{R}^n}
\frac{(x_i-y_i)\sum_{k=1}^{n}\partial g_k(\boldsymbol{y},t)/\partial y_k}
     {\lvert\boldsymbol{x}-\boldsymbol{y}\rvert^{\,n}}\,
\mathrm{d}\boldsymbol{y},
\end{equation}
the velocity equation collapses to the Cauchy diffusion equation and the pressure follows by direct
integration. This is the condition introduced by \citet{tha2023} and classified in
\citet{tha2026}, whose notation for the transverse field we retain; for periodic fields the
identity is understood through the corresponding Fourier Helmholtz projection, and in the notation of Section~\ref{sec:scheme} it states simply that the
transverse part of the convective field vanishes,
\begin{equation}
\label{eq:condP}
\Uc \;=\; \PT\boldsymbol{g} \;\equiv\; \boldsymbol{0},
\end{equation}
where $\PT$ is the Leray projection \citep{leray1934} onto divergence-free fields, defined in
Section~\ref{sec:scheme}. The two statements coincide because the Newtonian integral on the right
of \eqref{eq:cond} is the longitudinal part of $\boldsymbol{g}$ recovered from its divergence, so
that \eqref{eq:cond} asserts that $\boldsymbol{g}$ equals its own longitudinal part and
\eqref{eq:condP} that nothing transverse is left over. The convective field is then entirely
longitudinal and the pressure gradient balances it exactly. \citet{tha2026} classifies the periodic
fields satisfying \eqref{eq:cond} for $3\le n\le 8$, establishing existence at $n=3,4$ and
characterising the obstructions above.

The problem addressed here is the complementary one. When $\Uc\neq\boldsymbol{0}$ no such
reduction exists, and we ask whether the solution can nevertheless be constructed, by what
procedure, and within what range of the data.

\section{The construction in $n$ dimensions}
\label{sec:scheme}

Taking the divergence of \eqref{eq:ns} and using $\bnabla\bcdot\boldsymbol{v}=0$ gives the pressure
Poisson equation, whose role and pitfalls are discussed at length by \citet{gs1987},
\begin{equation}
\label{eq:ppe}
\Delta p = -\rho\,\bnabla\bcdot\boldsymbol{g}.
\end{equation}

Two standard notions organise what follows, and we state them explicitly because the argument turns
on the distinction between them. Any smooth periodic vector field $\boldsymbol{w}$ splits uniquely
into a gradient and a divergence-free remainder,
\begin{equation}
\label{eq:helm}
\boldsymbol{w} = \bnabla\phi + \boldsymbol{h},
\qquad \bnabla\bcdot\boldsymbol{h}=0,
\end{equation}
the \emph{Helmholtz decomposition}; $\bnabla\phi$ is the longitudinal part of $\boldsymbol{w}$ and
$\boldsymbol{h}$ its transverse part. In Fourier variables the split is immediate: the longitudinal
part of the mode $\hat{\boldsymbol{w}}_{\boldsymbol{k}}$ is its component along $\boldsymbol{k}$,
and the transverse part is what remains. The map returning the transverse part is the \emph{Leray
projection} \citep{leray1934}, written $\PT$. It removes from a field exactly what a pressure
gradient can supply.

Comparing \eqref{eq:ppe} with \eqref{eq:helm}, the pressure gradient is precisely the longitudinal
part of $\boldsymbol{g}$, and \eqref{eq:ns} is equivalent to
\begin{equation}
\label{eq:proj}
\frac{\partial\boldsymbol{v}}{\partial t} = \kappa\Delta\boldsymbol{v} - \PT\boldsymbol{g},
\end{equation}
with the pressure eliminated. The equivalence of \eqref{eq:cond} and \eqref{eq:condP} noted in
Section~\ref{sec:problem} is now transparent: the velocity satisfies pure diffusion precisely when
$\PT\boldsymbol{g}\equiv\boldsymbol{0}$.

Throughout, $\lVert\cdot\rVert$ denotes the cell-averaged $L^2$ norm,
$\lVert\boldsymbol{w}\rVert=\langle\lvert\boldsymbol{w}\rvert^2\rangle^{1/2}$, which by Parseval's
identity is the $\ell^2$ norm of the Fourier coefficients; the same norm is used for the successive
differences, the defect, and the divergences reported below.

Outside that class we iterate:
\begin{equation}
\label{eq:picard}
\boldsymbol{v}^{(m+1)}(t)
= e^{\kappa t\Delta}\boldsymbol{v}^{0}
- \int_0^t e^{\kappa(t-\tau)\Delta}\,
  \PT\bigl(\boldsymbol{v}^{(m)}\bcdot\bnabla\bigr)\boldsymbol{v}^{(m)}(\tau)\,\mathrm{d}\tau ,
\end{equation}
with $\boldsymbol{v}^{(1)}=e^{\kappa t\Delta}\boldsymbol{v}^{0}$. The reference velocity $v_r$
enters only through $\boldsymbol{v}^0$, so that $\boldsymbol{v}^{(m)}$ carries $v_r$ to the first
power in its leading term and to the $m$th in the term generated at step~$m$. Nothing in
\eqref{eq:picard} is dimensional. Every iterate is solenoidal by construction, since $\PT$ removes the longitudinal part
at each step and the heat semigroup preserves solenoidality; that property is what makes the
iterates candidate velocity fields, and Section~\ref{sec:3D} is in part about what happens when it
is lost. Once the velocity is known the pressure follows from \eqref{eq:ppe} by one division per
Fourier mode.

A family of initial fields carries the construction through every dimension. Writing indices
cyclically,
\begin{equation}
\label{eq:v0n}
\frac{v_i^0}{v_r} = \sin(\alpha x_i)\sin(\alpha x_{i-1})
      + \cos(\alpha x_i)\cos(\alpha x_{i+1}),
\qquad i=1,\dots,n .
\end{equation}
This field is solenoidal for every $n$: the two sums generated by $\partial_iv_i^0$ are carried into
one another by the reindexing $j=i-1$. It is also a Laplacian eigenfunction,
\begin{equation}
\label{eq:eig}
\Delta\boldsymbol{v}^0 = -2\alpha^2\boldsymbol{v}^0 ,
\end{equation}
with an eigenvalue \emph{independent of $n$}, so that the diffusive timescale does not shift between
dimensions and comparisons across $n$ are made at fixed viscous decay. Both properties have been
verified symbolically for $2\le n\le8$.

The case $n=2$ is degenerate and is set aside. There $\boldsymbol{g}^0$ vanishes identically for
\eqref{eq:v0n}, and more generally a two-dimensional field derived from a stream function that is a
Laplacian eigenfunction has a convective field that is a gradient. The argument is one line: with
$\boldsymbol{v}=(\partial_2\psi,-\partial_1\psi)$ the vorticity is $\omega=-\Delta\psi$, so
$\Delta\psi=-\lambda\psi$ gives $\omega=\lambda\psi$ and hence
$(\boldsymbol{v}\bcdot\bnabla)\omega=\lambda(\boldsymbol{v}\bcdot\bnabla)\psi=0$; since in two
dimensions $\bnabla\times\boldsymbol{g}=(\boldsymbol{v}\bcdot\bnabla)\omega$, the convective field
is curl-free and \eqref{eq:cond} holds automatically. The Taylor vortex is the canonical
instance. Leaving the solvable class in two dimensions therefore requires superposing two
eigenvalues, and the construction below is carried out for $n\ge3$.

\section{Three dimensions: the principal example}
\label{sec:3D}

At $n=3$ the family \eqref{eq:v0n} reads
\begin{align}
\label{eq:v0}
\frac{v_1^0}{v_r} &= \sin(\alpha x_1)\sin(\alpha x_3) + \cos(\alpha x_1)\cos(\alpha x_2),
\nonumber\\
\frac{v_2^0}{v_r} &= \sin(\alpha x_2)\sin(\alpha x_1) + \cos(\alpha x_2)\cos(\alpha x_3), \\
\frac{v_3^0}{v_r} &= \sin(\alpha x_3)\sin(\alpha x_2) + \cos(\alpha x_3)\cos(\alpha x_1),
\nonumber
\end{align}
with $\alpha=2\pi/L$ the wavenumber and $v_r$ the reference velocity. In the computations reported
below we take $\alpha=\pi$ and $v_r=1$, so that lengths are measured in units of $L/2$ and speeds in
units of $v_r$. Unlike the two-dimensional case, no superposition is needed: the
single-eigenvalue field already has
\begin{equation}
\label{eq:U0}
\Uc^0 = \PT\boldsymbol{g}^0 \neq \boldsymbol{0},
\end{equation}
so it lies outside the class of \citet{tha2026} and the iteration must be carried out.

Every Fourier mode of $\Uc^0$ carries $|\boldsymbol{k}|^2=6$ in units of $\alpha^2$, so $\Uc^0$ is
itself a Laplacian eigenfunction and the second iterate is available in closed form,
\begin{equation}
\label{eq:v2}
\boldsymbol{v}^{(2)} = \boldsymbol{v}^0e^{-2\alpha^2\kappa t}
-\Uc^0\,\frac{e^{-4\alpha^2\kappa t}-e^{-6\alpha^2\kappa t}}{2\alpha^2\kappa},
\end{equation}
in which $\Uc^0$ is quadratic in $v_r$ and carries a factor $\alpha$, so that the second term is a
velocity multiplied by $\alpha v_r/(\alpha^2\kappa)$---that is, by a Reynolds number.
The third iterate is where the construction ceases to be elementary. Forming
$(\boldsymbol{v}^{(2)}\bcdot\bnabla)\boldsymbol{v}^{(2)}$ produces two new spatial coefficients
carrying the time factors $\zeta_1\zeta_2$ and $\zeta_2^2$, where $\zeta_1=e^{-2\alpha^2\kappa t}$
and $\zeta_2$ is the bracket of \eqref{eq:v2}; the coefficients $\boldsymbol{\alpha}_1$ and
$\boldsymbol{\alpha}_2$ below are cubic and quartic in $v_r$ respectively:
\begin{equation}
\label{eq:alpha}
\alpha_{1,i} = -\sum_j \mathcal{U}_j^0\frac{\partial v_i^0}{\partial x_j}
               -\sum_j v_j^0\frac{\partial \mathcal{U}_i^0}{\partial x_j},
\qquad
\alpha_{2,i} = \sum_j \mathcal{U}_j^0\frac{\partial \mathcal{U}_i^0}{\partial x_j},
\end{equation}
\begin{equation}
\label{eq:v3}
\boldsymbol{v}^{(3)} = \boldsymbol{v}^{(2)}
- \int_0^t e^{\kappa(t-\tau)\Delta}\,
  \PT\bigl[\boldsymbol{\alpha}_1\zeta_1\zeta_2+\boldsymbol{\alpha}_2\zeta_2^2\bigr](\tau)\,
  \mathrm{d}\tau .
\end{equation}
Neither $\boldsymbol{\alpha}_1$ nor $\boldsymbol{\alpha}_2$ is a Laplacian eigenfunction; each
spreads over several wavenumbers and no closed form of the kind \eqref{eq:v2} exists. The projection has of
course already acted once, in forming $\Uc^0=\PT\boldsymbol{g}^0$ from a convective field whose
divergence is not zero. What is new at the third step is that $\boldsymbol{\alpha}_1$ and
$\boldsymbol{\alpha}_2$ are themselves not transverse, so that they cannot be returned to the
velocity equation as they stand; this is the first place at which the projection can be omitted by
oversight, a point we return to below.

Beyond this the expressions cease to be writable. The number of distinct Fourier modes carried by
the iterates is
\begin{equation}
\label{eq:modes}
12,\quad 36,\quad 168,\quad 684,\quad 744
\qquad\text{for}\qquad
\boldsymbol{v}^{(1)},\dots,\boldsymbol{v}^{(5)} ,
\end{equation}
so that $\boldsymbol{v}^{(4)}$ carries four times the spatial content of $\boldsymbol{v}^{(3)}$ and
would require some seven hundred trigonometric products, each with its own combination of
exponential time factors. We therefore give $\boldsymbol{v}^{(3)}$ as the last iterate in analytic
form and compute the remainder spectrally; \eqref{eq:modes} is itself a useful measure of the cost,
since the growth is close to geometric until the truncation begins to bite at the fifth step.

\subsection{How special is this field?}

Solenoidal fields sharing the eigenvalue $-2\alpha^2$ form a real vector space of dimension $24$:
twelve wavevectors have $|\boldsymbol{k}|^2=2$, hence six conjugate pairs, each carrying two complex
amplitudes orthogonal to $\boldsymbol{k}$. What distinguishes \eqref{eq:v0} is the spectral content
of $\Uc^0$. For a generic member of the same eigenspace $\Uc^0$ contains the three eigenvalues
$|\boldsymbol{k}|^2\in\{2,4,6\}$---we find this in every one of forty randomly drawn members---while
for \eqref{eq:v0} it contains $|\boldsymbol{k}|^2=6$ alone. That monochromatic property is what
makes \eqref{eq:v2} available in closed form. Convergence does not depend on it.

\subsection{Convergence, and the solution}
\label{sec:converge}

\begin{figure}[ht]
\centering
\includegraphics[width=0.84\textwidth]{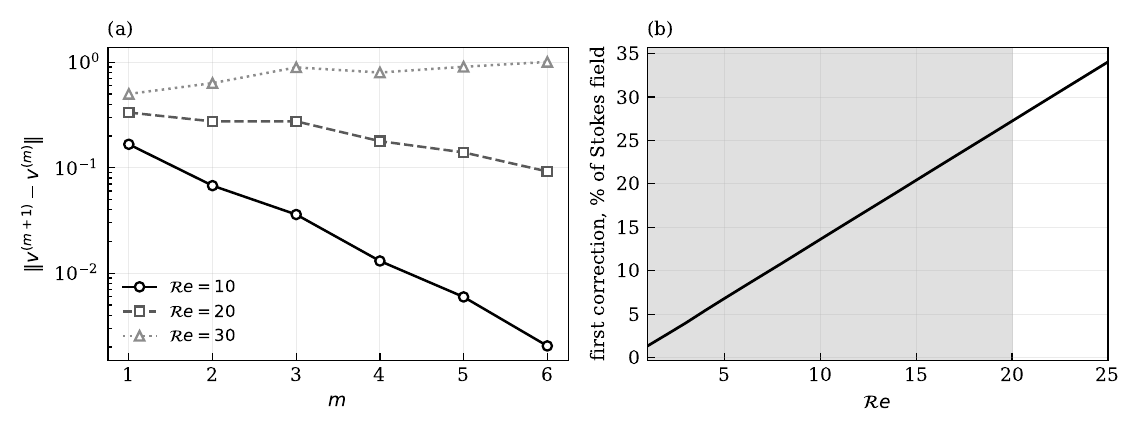}
\caption{(a)~Successive differences
$\lVert\boldsymbol{v}^{(m+1)}-\boldsymbol{v}^{(m)}\rVert$ against iterate order, logarithmic scale,
for the three-dimensional field \eqref{eq:v0}: the sequence contracts at $\mathcal{R}e=10$ and
$\mathcal{R}e=20$ and diverges at and above $\mathcal{R}e=30$, with the factors
of~\eqref{eq:factors}. (b)~The first inertial correction relative to the Stokes field, measured as
$\max_t\lVert\boldsymbol{v}^{(2)}-\boldsymbol{v}^{(1)}\rVert/\max_t\lVert\boldsymbol{v}^{(1)}\rVert$
from the closed form \eqref{eq:v2}; the dependence is linear in $\mathcal{R}e$ and passes ten per
cent near $\mathcal{R}e\approx7.5$, and the shaded band is the convergent range. Panel~(b) is
discussed in Sec.~\ref{sec:practice}.}
\label{fig:conv}
\end{figure}

\begin{table}[ht]
\centering
\begin{tabular}{lrr}
\toprule
$m\to m+1$ & $\mathcal{R}e=10$ & $\mathcal{R}e=20$ \\
\midrule
$1\to2$ & $1.667\times10^{-1}$        & $3.333\times10^{-1}$ \\
$2\to3$ & $6.758\times10^{-2}$ (0.406) & $2.747\times10^{-1}$ (0.824) \\
$3\to4$ & $3.599\times10^{-2}$ (0.533) & $2.749\times10^{-1}$ (1.001) \\
$4\to5$ & $1.305\times10^{-2}$ (0.363) & $1.784\times10^{-1}$ (0.649) \\
$5\to6$ & $5.954\times10^{-3}$ (0.456) & $1.396\times10^{-1}$ (0.783) \\
$6\to7$ & $2.044\times10^{-3}$ (0.343) & $9.212\times10^{-2}$ (0.660) \\
\bottomrule
\end{tabular}
\caption{Successive differences $\lVert\boldsymbol{v}^{(m+1)}-\boldsymbol{v}^{(m)}\rVert$,
maximised over $t\in[0,6]$, with the ratio to the preceding difference in parentheses. The maximum
is attained well before $t=4$ in every case, so restricting to $[0,4]$ changes no entry.}
\label{tab:contract}
\end{table}

At $\mathcal{R}e=10$ the differences fall by a factor of $82$ in six steps with ratios between
$0.34$ and $0.53$: over the orders examined the computed iteration contracts steadily and
converges numerically towards a fixed point of \eqref{eq:picard}. We record this as an observation
about the computation rather than as a proof, since a finite sequence of decreasing differences,
however clear, does not establish that the map of \eqref{eq:picard} is a contraction on any
particular function space; that would require the mapping and contraction estimates of
\citet{fk1964} and \citet{kato1984} rather than a numerical trend. A fixed point of \eqref{eq:picard} is a mild solution of \eqref{eq:ns}, and we
verify this directly. Applying the right-hand side of \eqref{eq:picard} to the final iterate and
measuring how far the result moves gives, with nine iterates,
\begin{equation}
\label{eq:resid}
\bigl\lVert\text{defect}\bigr\rVert = 1.004\times10^{-4},
\qquad \text{relative to } \lVert\boldsymbol{v}\rVert:\ 8.20\times10^{-5}.
\end{equation}
The constructed field is a fixed point of \eqref{eq:picard} to about one part in twelve thousand,
for an initial condition that does not satisfy \eqref{eq:cond}. This is the principal result of the
paper. It should be read for exactly what it measures: \eqref{eq:resid} is the defect of the
\emph{discretised} map, and the quadrature and truncation errors of Section~\ref{sec:repro} bound
the distance to the continuous problem separately. It is evidence that the iteration has reached
its own fixed point, not an error estimate against \eqref{eq:ns}.

Measuring an effective contraction factor as $(d_{m_{\max}}/d_1)^{1/(m_{\max}-1)}$ over five steps
gives
\begin{equation}
\label{eq:factors}
0.428,\quad 0.812,\quad 1.169,\quad 1.828,\quad 3.364
\qquad\text{at}\qquad
\mathcal{R}e=10,\,20,\,30,\,40,\,50 ,
\end{equation}
so the observed factor crosses unity between $\mathcal{R}e=20$ and $30$. At $\mathcal{R}e=20$ the
sequence contracts at $0.812$, slowly but usefully. This identifies the practical range over which
the iteration is usable for the initial field and the tests considered here,
$\mathcal{R}e\lesssim20$ with rapid convergence below about $15$; it is not a demonstrated
mathematical boundary lying between $20$ and $30$. The behaviour is consistent with the classical
small-data condition of \citet{fk1964} and \citet{kato1984}.

\paragraph{What this range means.} Throughout, $\mathcal{R}e$ denotes the reciprocal viscosity in
the normalisation used here, $\kappa=1/\mathcal{R}e$ with the field of \eqref{eq:v0} built at
wavenumber $\alpha=\pi$, so that the diffusive factor is $e^{-2\pi^2 t/\mathcal{R}e}$. Referred to
the peak speed $\lVert\boldsymbol{v}^0\rVert_\infty=\sqrt3$ and the length $1/\alpha$, the
conventional Reynolds number is
$\mathcal{R}e^{*}=\lVert\boldsymbol{v}^0\rVert_\infty\,\mathcal{R}e/\pi$, so that the values
$\mathcal{R}e=10,20,30$ correspond to $\mathcal{R}e^{*}\simeq5.5,\,11,\,17$. The construction is
therefore confined to the creeping and weakly inertial regime, and we make no claim about it
elsewhere.

This is a genuine restriction and it should be read as such; what the range does and does not
permit is taken up in Section~\ref{sec:practice}.

\subsection{Two invariants}
\label{sec:tests}

Two quantities keep the iterates admissible, and both are cheap enough to evaluate at every step.

\paragraph{Divergence of the returned source.} Only $\PT\boldsymbol{g}$ may be returned to the
velocity equation; the divergence of every field so returned must vanish to round-off. That the
test is not vacuous is shown by \citet{tha2016}, where the coefficients \eqref{eq:alpha} are
returned unprojected, on the stated ground that they are already divergence-free.
Table~\ref{tab:div} shows otherwise.

\begin{table}[ht]
\centering
\begin{tabular}{lr}
\toprule
field & $\lVert\bnabla\bcdot(\cdot)\rVert$ \\
\midrule
$\boldsymbol{v}^0$              & $0$ \\
$\Uc^0=\PT\boldsymbol{g}^0$     & $2\times10^{-15}$ \\
$\boldsymbol{g}^0$              & $6.044$ \\
$\boldsymbol{\alpha}_1$         & $107.4$ \\
$\boldsymbol{\alpha}_2$         & $154.6$ \\
$\PT\boldsymbol{\alpha}_1$      & $1\times10^{-14}$ \\
$\PT\boldsymbol{\alpha}_2$      & $1\times10^{-14}$ \\
\bottomrule
\end{tabular}
\caption{Divergence of the fields entering the third step. The initial field and the projected
sources are solenoidal to machine precision; the unprojected coefficients are not, by more than an
order of magnitude relative to the convective field itself.}
\label{tab:div}
\end{table}

\begin{figure}[ht]
\centering
\includegraphics[width=0.78\textwidth]{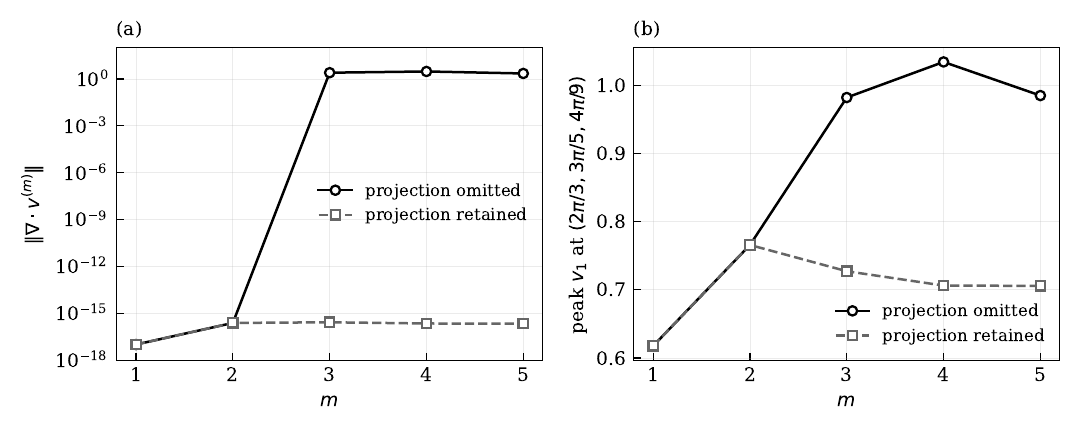}
\caption{The divergence test at $\mathcal{R}e=20$, logarithmic scale. Solid curves omit the
projection at the third and later steps; dashed curves retain it. With the projection
$\lVert\bnabla\bcdot\boldsymbol{v}^{(m)}\rVert$ remains at round-off, $2\times10^{-16}$, at every
order; without it the constraint is lost from the third iterate onwards, reaching $3.0$ by the
fourth---a separation of sixteen orders of magnitude at the step where the fault occurs.
Panel~(b) shows the corresponding velocity at $(2\pi/3,3\pi/5,4\pi/9)$.}
\label{fig:proj}
\end{figure}

Splitting by \eqref{eq:helm}, the longitudinal part carries $51.5\%$ of the energy of
$\boldsymbol{\alpha}_1$ and $41.8\%$ of that of $\boldsymbol{\alpha}_2$: roughly half of what is
returned belongs to the pressure. The mechanism deserves care, because the obvious guess is wrong.
The omitted longitudinal part does no work against the velocity---it is a gradient, and
$\langle\boldsymbol{v}^0\bcdot\bnabla\phi\rangle=0$ exactly, by orthogonality. The damage is done
through the constraint: once the unprojected source is integrated into the velocity the iterate
acquires a longitudinal component, is no longer divergence-free, and the pressure no longer removes
it, having been determined on the assumption that it was already absent. Figure~\ref{fig:proj}
shows both halves of this.

\paragraph{The energy inequality.} For an unforced viscous flow the cell-averaged kinetic energy
cannot increase. Table~\ref{tab:energy} reports its maximum over time for the projected iteration;
the initial value is $0.750$.

\begin{table}[ht]
\centering
\begin{tabular}{lrrr}
\toprule
& $\mathcal{R}e=20$ & $\mathcal{R}e=30$ & $\mathcal{R}e=50$ \\
\midrule
$\boldsymbol{v}^{(1)}$ & $0.7499$ & $0.7499$ & $0.7499$ \\
$\boldsymbol{v}^{(2)}$ & $0.7499$ & $0.7499$ & $0.7499$ \\
$\boldsymbol{v}^{(3)}$ & $0.7499$ & $0.7499$ & $1.633$ \\
$\boldsymbol{v}^{(4)}$ & $0.7499$ & $0.7499$ & $10.46$ \\
$\boldsymbol{v}^{(5)}$ & $0.7499$ & $0.7499$ & $543.4$ \\
\bottomrule
\end{tabular}
\caption{Maximum over $t$ of the cell-averaged kinetic energy, projected iteration. The initial
value is $0.750$. Within the convergent range no iterate exceeds it at any order; at
$\mathcal{R}e=50$ the iterates violate the energy inequality outright.}
\label{tab:energy}
\end{table}

At $\mathcal{R}e=50$ the fifth iterate reaches a cell energy of $543$ against an initial $0.750$,
with no forcing present. No solution of \eqref{eq:ns} can do this, so the growth is a property of
the expansion and not of the flow. The converse does not hold, and the limits of this diagnostic
should be stated plainly. The decay law applies to the exact unforced solution; a finite iterate of
\eqref{eq:picard} is driven by the projected convective field of the preceding iterate and is under
no obligation to obey it. An energy violation therefore identifies an inadmissible approximation,
but does not show that the iteration cannot converge, and an iterate that stays below the initial
energy is not thereby accurate: at $\mathcal{R}e=20$ and $30$ the energy sits at $0.7499$ at every
order whatever the quality of the quadrature, so it detects nothing in the range where the
construction is actually used. Of the two tests of this section only the divergence is sharp. The point generalises: truncated expansions are used to probe
singular behaviour, and an iterate that grows without bound invites the reading that the solution
does too. The present computation is a counterexample to that inference. The question of global
regularity for smooth data is stated precisely in \citet{fefferman}; expansions of this kind do not
bear on it unless they are first shown to converge.

\subsection{A note on the origin of the construction}

The construction developed here originates in \citet{tha2016}, from which three essential elements
are retained: the decomposition of \eqref{eq:ns} into viscous, inertial and applied contributions,
which also underlies the classification in \citet{tha2026}; the identification of $\Uc\equiv0$ as
the condition distinguishing the directly solvable problems from the remaining cases; and the
iterative construction itself. The first two iterates derived in \citet{tha2016} remain valid, and
\eqref{eq:v2} is exact.

The third step, however, is not correct: the solenoidal projection is omitted for the reason stated
above, and the growth reported there is a consequence of that omission rather than of the
Navier--Stokes dynamics. The calculations presented here correct and supersede those results. The
source of the discrepancy is precise: it first arises when the nonlinear source ceases to be a
Laplacian eigenfunction, which is also the first stage at which the projection removes a non-zero
component.

\section{Four to eight dimensions}
\label{sec:dims}

The family \eqref{eq:v0n} extends the three-dimensional field to every $n$, and for it the
quantities governing the construction are available in closed form. Direct computation gives
\begin{equation}
\label{eq:dimscaling}
\lVert\boldsymbol{g}^0\rVert = \frac{\sqrt{3n}}{4},
\qquad
\lVert\Uc^0\rVert = \sqrt{\frac{n}{6}},
\qquad
\frac{\lVert\Uc^0\rVert}{\lVert\boldsymbol{g}^0\rVert} = \frac{2\sqrt2}{3} \approx 0.9428 ,
\end{equation}
in units with $\alpha=v_r=1$, together with
\begin{equation}
\label{eq:longthird}
\frac{\lVert(I-\PT)\boldsymbol{g}^0\rVert}{\lVert\boldsymbol{g}^0\rVert}=\frac13 ,
\end{equation}
so that the convective field splits in energy as $8/9$ transverse to $1/9$ longitudinal. These hold
for every $n\ge3$ and are established in \citet{tha2026b}, where the mechanism is identified: the
convective field of \eqref{eq:v0n} is confined to the single spectral shell
$\lvert\boldsymbol{k}\rvert^2=6\alpha^2$, so that the Leray multiplier has constant denominator
across its entire support and the longitudinal energy follows from the divergence alone. We
reproduce the statement here because the construction of Section~\ref{sec:3D} depends on it, and
record in Table~\ref{tab:dims} the magnitudes as computed independently by the script of
Appendix~\ref{app:verify}.

\begin{table}[ht]
\centering
\begin{tabular}{crrr}
\toprule
$n$ & $\lVert\boldsymbol{g}^0\rVert$ & $\lVert\Uc^0\rVert$ & transverse fraction \\
\midrule
2 & $0$      & $0$      & --- \\
3 & $0.7500$ & $0.7071$ & $0.9428$ \\
4 & $0.8660$ & $0.8165$ & $0.9428$ \\
5 & $0.9682$ & $0.9129$ & $0.9428$ \\
6 & $1.0607$ & $1.0000$ & $0.9428$ \\
7 & $1.1456$ & $1.0801$ & $0.9428$ \\
8 & $1.2247$ & $1.1547$ & $0.9428$ \\
\bottomrule
\end{tabular}
\caption{The convective field of \eqref{eq:v0n} and its transverse part, computed for
$2\le n\le8$ and in agreement with \eqref{eq:dimscaling} throughout. The magnitude grows as
$\sqrt{n}$ but the transverse fraction is independent of dimension. At $n=2$ the convective field
vanishes identically and the member is a solvable-class field in the sense of \eqref{eq:cond}.}
\label{tab:dims}
\end{table}

Two consequences follow. The convective field grows only as $\sqrt{n}$, so higher dimensions are
not intrinsically more nonlinear for this family. And the fraction of it that the pressure cannot
absorb is the \emph{same} in every dimension, so the construction is needed to the same degree
throughout the range. The strong dimensional dependence that governs the classification in
\citet{tha2026}---existence at $n=3,4$, obstruction above---is absent here, which is to be expected:
that dependence concerns whether \eqref{eq:cond} can be satisfied at all, a question that does not
arise once the iteration is available.

\section{Use beyond benchmarking}
\label{sec:practice}

\subsection*{A closed-form inertial correction, and where it applies}

The iterates form an expansion in powers of the nonlinearity about the Stokes solution:
\eqref{eq:v2} is the first inertial correction in closed form, \eqref{eq:v3} the second, and so on.

The range established in Section~\ref{sec:converge}, $\mathcal{R}e^{*}\lesssim11$, is precisely the
range in which such expansions are used. Low-Reynolds-number regimes of this order occur, for
example, in lubrication, microfluidics, porous-media flow and small-vessel haemodynamics, although
the range spanned within each of those areas is wider and depends on geometry and working fluid: in journal bearings and thin-film flows inertia is customarily neglected outright; in
microfluidic devices channel dimensions of tens of microns place $\mathcal{R}e^{*}$ below ten for
most working fluids; Darcy and Brinkman descriptions of porous media assume creeping flow, with
inertial corrections entering through the Forchheimer term at the upper end of this range; and flow
in the smaller vessels of the circulation is likewise low-inertia. In each the recurring question is
the same---at what point does inertia cease to be negligible---and it is usually answered by
scaling estimate rather than by computation. These are cited to locate the regime, not as
applications: the present family is a periodic unforced flow on a cell, and none of these systems
is studied here. Its geometry, boundary conditions and forcing are all different, and we make no
claim that \eqref{eq:v0n} is representative of any of them.

Figure~\ref{fig:conv}(b) answers it directly for the present family. The first correction reaches
ten per cent of the Stokes field near $\mathcal{R}e\approx7.5$, that is
$\mathcal{R}e^{*}\approx4.1$, and twenty-seven per cent by $\mathcal{R}e=20$, growing linearly. The
correction is available in closed form, so the estimate is a computation rather than an
order-of-magnitude argument. We do not claim more: \eqref{eq:factors} is an observed rate, not a
bound, and converting it into an a~priori estimate of the number of terms required for a stated
accuracy would need the contraction estimates of \citet{fk1964} and \citet{kato1984}, which we do
not supply.

What the construction does not offer is access to the transitional or turbulent regimes. Beyond the
threshold the expansion diverges, and---as Section~\ref{sec:tests} shows through the energy
test---it does so in a way that carries no information about the flow. The restriction is therefore
sharp rather than gradual: within the range the iterates are usable and verifiable, and outside it
they are not to be read at all.

The projection $\PT$ in \eqref{eq:picard} is also the operation performed at every time step of a
fractional-step method \citep{chorin1968}: advance the velocity, then remove the longitudinal part.
Such methods dominate incompressible computation, and errors in the projection step are known to be
a recurring difficulty \citep{gs1987} and are quiet---they degrade the solution without announcing
themselves. The first invariant of Section~\ref{sec:tests} transfers unchanged and costs almost
nothing: monitoring $\lVert\bnabla\bcdot\boldsymbol{v}\rVert$ after each projection requires one
divergence evaluation per step, needs no reference solution, and separates a correct scheme from a
faulty one by fourteen orders of magnitude at the step where the fault occurs. Neither invariant is
novel in itself; what the present example supplies is a quantitative demonstration of how much they
catch, and how early.

\section{Reproducibility}
\label{sec:repro}

The principal numerical results reported here are produced by a single script, listed in Appendix~\ref{app:verify} and also supplied as an ancillary file: the divergences of Table~\ref{tab:div}, the longitudinal fractions, the spectral content of $\Uc^0$, and the peak velocities and cell energies of Tables~\ref{tab:contract} and~\ref{tab:energy}. It requires Python~3.8 or later and NumPy and reads no data files; run times depend on the hardware used.

The convolution integral is evaluated by trapezoidal quadrature, with half weights at both
endpoints; the time grid begins at $t=10^{-4}$ rather than $0$, so an interval of that length is
omitted from every convolution, which is below the resolution of every figure reported.
The fields are held as Fourier coefficients on the periodic cell, so the Leray projection is the
removal of the component along $\boldsymbol{k}$ and the divergence is
$\mathrm{i}\alpha\boldsymbol{k}\bcdot\hat{\boldsymbol{v}}_{\boldsymbol{k}}$; both are exact on the
represented modes, and the divergences of order $10^{-14}$ in Table~\ref{tab:div} are arithmetic
round-off. The only approximations are the trapezoidal quadrature of the time convolution in
\eqref{eq:picard} and the Fourier truncation.

Two resolution checks bound both, and the temporal one is reported in full because the convergence
of the quadrature is not incidental to the results. The effective contraction factor of
\eqref{eq:factors} at $\mathcal{R}e=20$, five iterates, against the number of time points:
\begin{equation}
\label{eq:tconv}
0.8238,\quad 0.8137,\quad 0.8118,\quad 0.8112
\qquad\text{at}\qquad
n_t=60,\,120,\,240,\,480 ,
\end{equation}
so the figures reported here, computed at $n_t=240$, are converged to within $10^{-3}$ and the
convergence is second order in the step, as trapezoidal quadrature requires. Enlarging the
truncation from $|k_i|\le6$ to $|k_i|\le8$ changes as little: at $\mathcal{R}e=20$ the cell energy
remains $0.7499$ at every order, and at $\mathcal{R}e=50$ the five orders give $0.7499$, $0.7499$,
$1.6327$, $10.4588$ and $544.4$ against $0.7499$, $0.7499$, $1.6327$, $10.4581$ and $543.4$ at the
smaller truncation. Neither approximation is limiting at the resolutions used.

\section{Concluding remarks}
\label{sec:conclusions}

For initial fields whose convective term is not a pure gradient, and which therefore fall outside
the directly solvable class, the projected iteration provides a constructive route to the solution
within its numerically observed convergence range. In three dimensions, at $\mathcal{R}e=10$, the
observed contraction factor is $0.43$, and the field the iteration converges to is a fixed point of
\eqref{eq:picard} to a relative defect of $8.2\times10^{-5}$.

The construction has a definite range of applicability. The observed contraction factor increases
with Reynolds number, reaching $0.812$ at $\mathcal{R}e=20$ and exceeding unity by
$\mathcal{R}e=30$, giving a practical range of $\mathcal{R}e\lesssim20$ for the initial field
considered here.

The construction itself is dimension-independent. For the family \eqref{eq:v0n}, the fraction of the
convective field that cannot be absorbed into the pressure is exactly $2\sqrt{2}/3$ for every
$n\ge3$, and the longitudinal fraction exactly $1/3$; the reason is that the convective field of
\eqref{eq:v0n} occupies a single spectral shell, $\lvert\boldsymbol{k}\rvert^2=6\alpha^2$. At
$n=2$ this fraction vanishes and no iteration is required. The two-dimensional case is exceptional
for a structural reason: no single-eigenvalue field in two dimensions leaves the directly solvable
class.

Of the two invariants examined, one is a sharp diagnostic for any iteration of this type and one is
not. The divergence of every field returned to the velocity equation separates a correct scheme
from a faulty one by sixteen orders of magnitude at the step where the fault occurs. The
cell-averaged kinetic energy is far blunter: it registers the gross failure at
$\mathcal{R}e=50$ but is silent throughout the range in which the construction is used. More
broadly, apparent growth in a truncated expansion is, by itself, a property of the expansion rather
than evidence of growth in the underlying Navier--Stokes solution. Convergence must first be
established before dynamical conclusions can be drawn from the expansion.

\appendix
\section{The verification script}
\label{app:verify}

The script below reproduces Tables~\ref{tab:contract}--\ref{tab:dims}, the longitudinal fractions,
the spectral content of $\Uc^0$, the residual \eqref{eq:resid} and the refinement checks
\eqref{eq:tconv} of Section~\ref{sec:repro}. It requires Python~3.8 or later and NumPy, reads no
data files, and is run by \texttt{python3 verify\_proj.py}. The convective derivative is evaluated
spectrally on an alias-free grid and truncated back to $|k_i|\le K$, which is identical to a direct
sum over mode pairs but fast enough that the nine-iterate residual and the sweep over
$\mathcal{R}e$ are reachable; seven iterates at $n_t=240$ take about sixteen seconds. Both
quadrature rules are retained so that \eqref{eq:tconv} can be reproduced.

\begingroup\scriptsize
\begin{verbatim}
#!/usr/bin/env python3
"""Corrected and complete verification for NSPROJ.

Reproduces every numerical result claimed in the paper: Tables I-IV, the
longitudinal fractions, the spectral content of U^0, the residual of Eq.(16),
and the temporal/truncation refinement checks of Sec. VI.

Two substantive changes from the submitted script:
  * the time-convolution quadrature really is trapezoidal;
  * (A.grad)B is evaluated spectrally (identical result, ~100x faster), which
    is what makes the nine-iterate residual and the Re sweep reproducible.
"""

import itertools, math, numpy as np

class Space:
    def __init__(self, K=6, n=3, M=None):
        self.K, self.n = K, n
        self.M = M or (4*K + 2)                 # alias-free: products reach 2K
        self.ks = list(itertools.product(range(-K, K+1), repeat=n))
        self.idx = {k: i for i, k in enumerate(self.ks)}
        self.N = len(self.ks)
        self.lam = np.array([sum(t*t for t in k)*math.pi**2 for k in self.ks])
        self.kk = np.array(self.ks).T                      # (n, N)
        M = self.M
        self.sl = tuple(np.array([k[d] % M for k in self.ks]) for d in range(n))

    def zero(self): return np.zeros((self.n, self.N), complex)

    def to_grid(self, f):
        M, n = self.M, self.n
        G = np.zeros((M,)*n, complex)
        np.add.at(G, self.sl, f)
        return np.fft.ifftn(G)*M**n

    def from_grid(self, G):
        M, n = self.M, self.n
        H = np.fft.fftn(G)/M**n
        return H[self.sl]

    def grad_dot(self, A, B):
        """(A.grad)B, truncated to |k_i|<=K  (identical to the original)"""
        n = self.n
        Ag = [self.to_grid(A[j]) for j in range(n)]
        out = self.zero()
        for i in range(n):
            acc = np.zeros((self.M,)*n, complex)
            for j in range(n):
                acc += Ag[j]*self.to_grid(1j*math.pi*self.kk[j]*B[i])
            out[i] = self.from_grid(acc)
        return out

    def transverse(self, G):
        k2 = np.sum(self.kk**2, axis=0).astype(float)
        k2s = np.where(k2 == 0, 1.0, k2)
        kd = np.sum(self.kk*G, axis=0)/k2s
        out = G - self.kk*kd
        out[:, k2 == 0] = G[:, k2 == 0]
        return out

    def divnorm(self, F):
        return float(np.sqrt(np.sum(np.abs(np.sum(1j*math.pi*self.kk*F, axis=0))**2)))

    def v0(self):
        n = self.n
        f = self.zero()
        def put(c, co, k):
            f[c][self.idx[k]] += co/2
            f[c][self.idx[tuple(-t for t in k)]] += co/2
        for i in range(n):
            e = lambda d: tuple(1 if t == d else 0 for t in range(n))
            a, b, c = e(i), e((i-1) % n), e((i+1) % n)
            put(i, 0.5, tuple(a[d]-b[d] for d in range(n)))
            put(i,-0.5, tuple(a[d]+b[d] for d in range(n)))
            put(i, 0.5, tuple(a[d]-c[d] for d in range(n)))
            put(i, 0.5, tuple(a[d]+c[d] for d in range(n)))
        return f

def energy(vh): return 0.5*float(np.sum(np.abs(vh)**2).real)
def norm(F):    return float(np.sqrt(np.sum(np.abs(F)**2)).real)

def picard(S, Re, tgrid, orders, rule='trap'):
    kap = 1.0/Re; nt = len(tgrid); v0 = S.v0()
    E = np.exp(-np.outer(tgrid, kap*S.lam))
    seq = [np.array([v0*E[it] for it in range(nt)])]
    for _ in range(orders-1):
        gT = np.array([S.transverse(S.grad_dot(seq[-1][it], seq[-1][it]))
                       for it in range(nt)])
        vn = np.zeros((nt, S.n, S.N), complex)
        for it, t in enumerate(tgrid):
            vn[it] = v0*E[it]
            if it > 0:
                tau = tgrid[:it+1]
                if it > 1:
                    w = np.gradient(tau).astype(float)
                    if rule == 'trap': w[0] /= 2.0; w[-1] /= 2.0
                else:
                    w = np.array([tau[1]-tau[0]]*2)/2
                ker = np.exp(-np.outer(t-tau, kap*S.lam))
                vn[it] -= np.tensordot(w, gT[:it+1]*ker[:,None,:],
                                       axes=(0,0))
        seq.append(vn)
    return seq



def diffs(seq, mask):
    return [max(norm(seq[m+1][it]-seq[m][it])
                for it in np.where(mask)[0]) for m in range(len(seq)-1)]
def factor(d):  return (d[-1]/d[0])**(1.0/(len(d)-1))

S = Space(K=6, n=3)
v0 = S.v0(); g0 = S.grad_dot(v0, v0); U0 = S.transverse(g0)
a1 = -S.grad_dot(U0, v0) - S.grad_dot(v0, U0); a2 = S.grad_dot(U0, U0)

print("TABLE II  divergence of the fields entering the third step", flush=True)
for nm, F in (("v^0",v0),("U^0",U0),("g^0",g0),("alpha_1",a1),("alpha_2",a2),
              ("P alpha_1",S.transverse(a1)),("P alpha_2",S.transverse(a2))):
    print(f"   {nm:10s} ||div|| = {S.divnorm(F):.4f}", flush=True)
print(f"   ratio ||div a1||/||div g0|| = "
      f"{S.divnorm(a1)/S.divnorm(g0):.1f}", flush=True)

tg = np.linspace(1e-4, 6, 240)
m4 = tg <= 4.0          # Table I caption says max over t in [0,4]
m6 = np.ones_like(tg, bool)

print("\nTABLE I  successive differences, seven iterates", flush=True)
for Re in (10.0, 20.0):
    for rule in ('grad','trap'):
        seq = picard(S, Re, tg, 7, rule)
        d4, d6 = diffs(seq, m4), diffs(seq, m6)
        lab = 'np.gradient' if rule=='grad' else 'trapezoid  '
        print(f"   Re={Re:g} {lab} t<=4: "
              + " ".join(f"{x:.4e}" for x in d4), flush=True)
        print(f"   {'':17s}       t<=6: "
              + " ".join(f"{x:.4e}" for x in d6), flush=True)

print("\nEQ.(17)  effective contraction factor, five iterates, t<=6", flush=True)
print(f"   {'Re':>4} {'np.gradient':>12} {'trapezoid':>12}", flush=True)
for Re in (10.0,20.0,30.0,40.0,50.0):
    f = {}
    for rule in ('grad','trap'):
        f[rule] = factor(diffs(picard(S, Re, tg, 5, rule), m6))
    print(f"   {Re:4.0f} {f['grad']:12.4f} {f['trap']:12.4f}", flush=True)

print("\nTABLE III  max cell energy, trapezoid, five iterates", flush=True)
for Re in (20.0,30.0,50.0):
    seq = picard(S, Re, tg, 5, 'trap')
    en = [max(energy(v[it]) for it in range(len(tg))) for v in seq]
    print(f"   Re={Re:g}: " + " ".join(f"{e:9.4f}" for e in en), flush=True)

print("\nEQ.(16)  residual at Re=10, nine iterates, trapezoid", flush=True)
seq = picard(S, 10.0, tg, 10, 'trap')
dfct = max(norm(seq[-1][it]-seq[-2][it]) for it in range(len(tg)))
vn   = max(norm(seq[-1][it]) for it in range(len(tg)))
print(f"   ||defect|| = {dfct:.3e}   relative = {dfct/vn:.2e}", flush=True)

print("\nSEC.VI  temporal refinement, Re=20, five iterates", flush=True)
for nt in (60,120,240,480):
    t2 = np.linspace(1e-4,6,nt); mm = np.ones_like(t2,bool)
    r = {ru: factor(diffs(picard(S,20.0,t2,5,ru), mm))
         for ru in ('grad','trap')}
    print(f"   nt={nt:4d}  np.gradient {r['grad']:.4f}"
          f"   trapezoid {r['trap']:.4f}", flush=True)

print("\nSEC.VI  truncation check, Re=20 and 50, trapezoid, five iterates", flush=True)
for K in (6,8):
    SK = Space(K=K, n=3)
    for Re in (20.0,50.0):
        seq = picard(SK, Re, tg, 5, 'trap')
        en = [max(energy(v[it]) for it in range(len(tg))) for v in seq]
        print(f"   |k|<={K}, Re={Re:g}: "
              + " ".join(f"{e:9.4f}" for e in en), flush=True)
print("\nDone.", flush=True)
\end{verbatim}
\endgroup

\section*{Declarations}

\noindent\textbf{Data availability.} No experimental or measured datasets were generated or
analysed. All results are reproducible from the accompanying script.

\medskip
\noindent\textbf{Declaration of AI use}
A large language model (Claude, Anthropic) was used as an assistant for LaTeX preparation and running the symbolic and numerical
verification scripts. The mathematical results, their proofs, and the interpretation of all computations are the author's own.

\medskip
\noindent\textbf{Funding.} No funding was received for conducting this study.

\medskip
\noindent\textbf{Competing interests.} The author has no competing interests to declare.

\end{document}